\documentclass[12pt]{article}

\usepackage{newtxtext,newtxmath}

\usepackage{graphicx}

\usepackage[letterpaper,margin=1in]{geometry}

\renewenvironment{abstract}
	{\quotation}
	{\endquotation}

\date{}

\makeatletter
\renewcommand{\fnum@figure}{\textbf{Figure \thefigure}}
\renewcommand{\fnum@table}{\textbf{Table \thetable}}
\makeatother

\usepackage{scicite}

\usepackage{url}

\def\scititle{
	First Photon Machine Learning
}
\title{\bfseries \boldmath \scititle}

\author{
	Lili Li$^{1,2}$,
	Santosh Kumar$^{1,2}$,
	Malvika Garikapati$^{1,2,3}$
        Yu-Ping Huang$^{1,2,3\ast}$\and
	\small$^{1}$Department of Physics, Stevens Institute of Technology, Hoboken, New Jersey 07030, USA.\and
	\small$^{2}$Center for Quantum Science and Engineering, Stevens Institute of Technology, Hoboken, NJ, 07030, USA.\and
        \small$^{3}$Quantum Computing Inc., Hoboken, New Jersey, 07030, USA.\and
	\small$^\ast$Corresponding author. Email: yhuang5@stevens.edu
}

\begin{document} 

\maketitle

\begin{abstract} \bfseries \boldmath
Quantum techniques are expected to revolutionize how information is acquired, exchanged, and processed. Yet it has been a challenge to realize and measure their values in practical settings. We present first photon machine learning as a new paradigm of neural networks and establish the first unambiguous advantage of quantum effects for artificial intelligence. By extending the physics behind the double-slit experiment for quantum particles to a many-slit version, our experiment finds that a single photon can perform image recognition at around $30\%$ fidelity, which beats by a large margin the theoretical limit of what a similar classical system can possibly achieve (about 24\%). In this experiment, the entire neural network is implemented in sub-attojoule optics and the equivalent per-calculation energy cost is below $10^{-24}$ joule, highlighting the prospects of quantum optical machine learning for unparalleled advantages in speed, capacity, and energy efficiency.
\end{abstract}

\noindent
\section{INTRODUCTION}

Artificial intelligence (AI) \cite{stokel2023chatgpt,gurney2018introduction,Mocanu2018,Shastri2021,doi:10.1126/science.abn7293,Cerezo2022,RevModPhys.91.045002} and quantum information science (QIS) \cite{vedral2006introduction,Biamonte2017,Ladd2010,Liu2019,nielsen2010quantum,Peng:20} are at the very front of information technology. Both are pursued intensively by academia and industry, fueled by the unprecedented information capabilities each promises. Yet their development trajectories have been on quite different paths. AI is crafted on highly parallel and massive data processing using integrated digital circuits, and has quickly flourished in many application areas thanks to the rapid progress in semiconductor manufacturing and the availability of big data \cite{balcerzak2024introduction}. However, its future is shadowed by the exceeding energy consumption and data sizes required for training \cite{OUNIFI2022101967,GARCIAMARTIN201975}. Almost on the opposite, QIS can be extremely efficient in data processing and consume orders of magnitude less energy \cite{Maring2024,Liu2021,watrous2018theory,Dunjko_2018,doi:10.1126/science.1142892}. However, QIS devices and systems are hard to scale up, as the manufacturing complexity and operating overhead of quantum devices are much higher than digital circuits \cite{hayashi2016quantum}. 

Hence, it is natural to ask this question: can QIS and AI complement each other and work together to lay new grounds for information processing \cite{Shastri2021, Biamonte2017,dunjko2020non}? On the one hand, QIS has the potential to significantly increase the energy and data efficiency for AI \cite{PhysRevA.107.010101,Blais2020}. On the other hand, AI can assist QIS in making scalable quantum devices and systems for practical applications \cite{PhysRevA.107.010101}. This prospective has spurred lots of studies in this junction, with proposals on quantum convolutional neural networks \cite{Cong2019}, quantum associative memories \cite{PhysRevLett.130.190602,VENTURA2000273}, quantum-enhanced reinforcement learning \cite{PRXQuantum.2.010328,PhysRevLett.117.130501}, quantum variational algorithms \cite{Cerezo2021,Moll_2018}, and so on. Yet, despite some credible arguments, hitherto there has not been a convincing experiment proof that QIS does give an edge to AI \cite{vedral2006introduction,Liu2019}. 

In this paper, we present, for the first time, an unambiguous experimental evidence that quantum effects can indeed elevate machine learning above the performance ceiling allowed by any classical means. This is in contrast to previous demonstrations where the insertion of quantum elements in a neural network seems to somewhat improve its performance, but it is unclear if such improvement is material or can be achieved alternatively by better training \cite{Cerezo2022}. Rather, here we show that, for the same problem and using the same resources, even imperfect experimental results from an under-optimized quantum setup can already beat the theoretical performance of an optimal classical counterpart by a significant margin.

We would like to point out that a benchmark against a known performance ceiling established by rigorous mathematical proofs is difficult, but of both essential and practical importance. A good example is with quantum computing, where many claims of ``quantum supremacy'' in time-to-solution have been subsequently proven questionable or controversial \cite{Arute2019,10.1145/3458817.3487399,doi:10.1126/sciadv.abl9236,mccormick2022race,PhysRevLett.129.090502,cho2022ordinary}. This is mainly because those claims were made upon benchmarking against the best known classical solvers, not a proven theoretical limit. Beating a known best does not necessarily prove a quantum advantage, as better classical methods might be developed in the future. The establishment of quantum advantages in machine learning is even more complex, as neural networks themselves are usually intricate and inexplicable. 

Our demonstration focuses on the quantum superposition principle \cite{feynman2015feynman,Bian2023}, a foundation of quantum mechanics first observed in an electron-beam double slit experiment \cite{THOMSON1927}, where a single electrons can simultaneously take two paths to produce self interference \cite{PhysRevLett.116.253601,Steeds_2003,Kim2023}, akin to waves. This then-shocking observation of wave-particle duality \cite{selleri1992wave,PhysRevX.11.031041} has historically played the key role in the discovery and development of quantum mechanics \cite{griffiths2018introduction}. Here we expand the double slits to hundreds of ``slits'' and show in experiment that by having a single photon simultaneously interacting with many pixels of an image, pattern recognition can be achieved at a fidelity higher than what's theoretically possible in classical mechanics where such superposition is not allowed. Our experiment thus constitutes a clear and indisputable demonstration of quantum advantages in AI. 

Meanwhile, our experiment sets a new record for the energy efficiency of neural networks. All information is encoded on few photons, the entire neural network is implemented in the quantum optics domain, and the decision is made upon the detection of a single photon, whose energy is merely $10^{-19}$ joule. The equivalent energy consumption per calculation is much less than $10^{-24}$ joule, or about $10^{-6}$ photons, which is orders of magnitude lower than the state of the art \cite{Wang2022}. This highlights the prospect of the future AI systems built partially upon quantum optics, which, among other benefits, promises to address the anticipated electrical power crisis associated with quick and wide adoption of AI technology.

\section{Model}

Our experiment is inspired by an elegant proposal by Fischbacher and Sbaiz \cite{fischbacher2020single}. It takes on a common benchmark problem of Modified National Institute of Standards and Technology (MNIST) and fashion-MNIST pattern recognition \cite{deng2012mnist, xiao2017fashionmnistnovelimagedataset}, and considers an interesting task of making a decision after receiving a single quantum of signal. Through rigorous math derivations, it finds the optimum ``classical'' classifier for this task and calculated the highest possible accuracy to be 22.96\%. In contrast, a classifier utilizing quantum superposition can theoretically achieve an accuracy above 41.27\%, although no experimental implementation was proposed. 

\begin{figure} 
	\centering
	\includegraphics[width=0.8\textwidth]{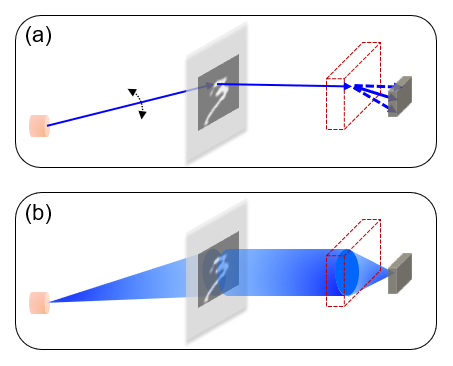} 

	\caption{\textbf{The conceptual design of CC (a) and QC (b).}
		In CC, single photons are allowed to sequentially (but not simultaneously) probe the pixels on a MNIST pattern using a beam rotator. In QC, they can be prepared in a quantum superposition state to probe the pixels simultaneously. After the pattern, the photons in both cases can pass through a unitary transformation circuit, represented by a dashed box, and detected by a single photon detector or a photon detector array.}
	\label{Fig.CM} 
\end{figure}

Figure~\ref{Fig.CM} illustrates the concept of our experiment. It deals with image classification, where each image is a standard binary pattern with $28\times28$ pixels from the MNIST handwritten digit set. The patterns are encoded as the spatial modes of a few-photon coherent signal through a binary spatial modulator with on/off pixels. The encoded signal is then measured with a single-photon sensitive device that can extract its spatial mode information, either directly with a single-photon detector array or by applying unitary transformation before the detection. For each pattern, only one photon-detection event is allowed, where its digit class (i.e., label) must be inferred after the first photon is detected, hence the name of first photon machine learning. 

Our goal is to use this paradigm to study and quantify quantum effects for machine learning. To this end, we construct two neural networks, a classical classifier (CC) and a quantum classifier (QC), as shown in \ref{Fig.CM}(a) and (b), respectively. For a fair comparison, the two use the same resources to recognize the same patterns under the same constraint: the classification is made upon the detection of the very first photon. Both can choose how the photons are measured and how the measurement results are processed. The only difference, however, is that the effect of quantum superposition is utilized in QC, but not permitted in CC.

For CC, Fischbacher and Sbaiz proved that the optimum classifier is realized by first identifying the pixel from which the detected photon comes, and then determining for each pixel what the most likely label is \cite{fischbacher2020single}. Here, the most likely label is the one containing the highest number of patterns having that particular pixel on. The classification fidelity obtained this way, which is the average accuracy for all patterns of each label, gives the accuracy threshold. No method, including redistributing light intensity among pixels, can help exceed this threshold. This can be understood intuitively through the Bayesian analysis: when there is only one feature detected, without any prior information, the best guess is by finding out which label can give this feature. For the whole MNIST handwritten data set, the accuracy threshold is $22.96\%$, which is significantly higher than that by random guessing ($10\%$). 

For the quantum classifier, only one photon can be detected, too. However, the photon could have interacted with multiple pixels, just like how an electron can take both paths in the double-slit experiment \cite{feynman2015feynman,Steeds_2003}. In addition, the photon can be measured in certain superposition states over multiple pixels, realizable by applying unitary transformation to it before the detection. Note that any transformation is permitted in the classical case, but it won't help increase the fidelity \cite{fischbacher2020single}. In this work, we propose and demonstrate that a simple projection of the encoded photons onto a set of Hermite Gaussian spatial modes is sufficient to beat the classical fidelity threshold.

In the following we use some simplified math to further illustrate the distinction between CC and QC. For both, a weak coherent laser beam on a few-photon level in a Gaussian spatial mode $\mathrm{G}(x,y)$ is used to illuminate the encoding modulator displaying the binary mask $M^{c}_{j}(x,y)$ of a MNIST digit pattern $j$ of Label $c$. The only difference is that in the classical case, the ``which way'' information is revealed, so that the transmitted photons is in an incoherent mixture state:
\begin{equation}
\label{eq1}
    P^c_{j}\sim\int \mathrm{G}^2(x,y)M^{c}_{j}(x,y)|x,y\rangle\langle x,y| dx dy 
\end{equation}
where we have used $M^{c}_{j}(x,y)^2= M^{c}_{j}(x,y)$ as it is a binary function. In the quantum case, on the other hand, a photon can simultaneously take multiple paths without knowing which (even in principle), so that the transmitted photon is in a pure state
\begin{equation}
\label{eq2}
   Q^c_{j}\sim\int \Psi^c_j(x,y) |x,y\rangle\langle x',y'| \Psi^{c}_j(x',y') dx\, dy\,dx'\,dy' 
\end{equation}
with $\Psi^c_j(x,y)=\mathrm{G}(x,y) M^{c}_{j}(x,y)$. 

The physics behind QC is thus identical to that behind the double-slit experiment \cite{feynman2015feynman}, where a quantum particle can go through two slits at the same time and self interfere on a screen. If a detector is placed behind the slits to measure which path the particle has taken, the interference pattern will disappear and reduce to the classical case. In our case, each pixel can be viewed as a slit, so that there are many ``slits'' to transmit the photons in the beam. When a photon is detected, there is no way to tell which ``slit'' it has taken, so that quantum mechanically, it has taken multiple paths coherently, giving nonzero off-diagonal terms in density matrix~(\ref{eq2}). On the other hand, for the classical classifier, the ``which way'' information for photons is revealed, causing their quantum state to collapse onto a mixture state in Eq.~(\ref{eq1}). 

For both classifiers, the photons can either be directly detected by a single-photon detector (array) or first undergo a unitary transformation before detection. The direct measurement gives the pixel location information, while the later projects the detected photons onto certain spatial modes. In this experiment, we consider the simplest case where the photons are projected onto 10 spatial modes. Each mode corresponds to a digit class, so that upon photon detection in mode $j$, the image is assigned to the corresponding Label $j$. In this way, no further data processing is required, and the entire neural network is implemented in the optics domain.

\section{EXPERIMENTAL SETUP}

\begin{figure} 
	\centering
	\includegraphics[width=0.9\textwidth]{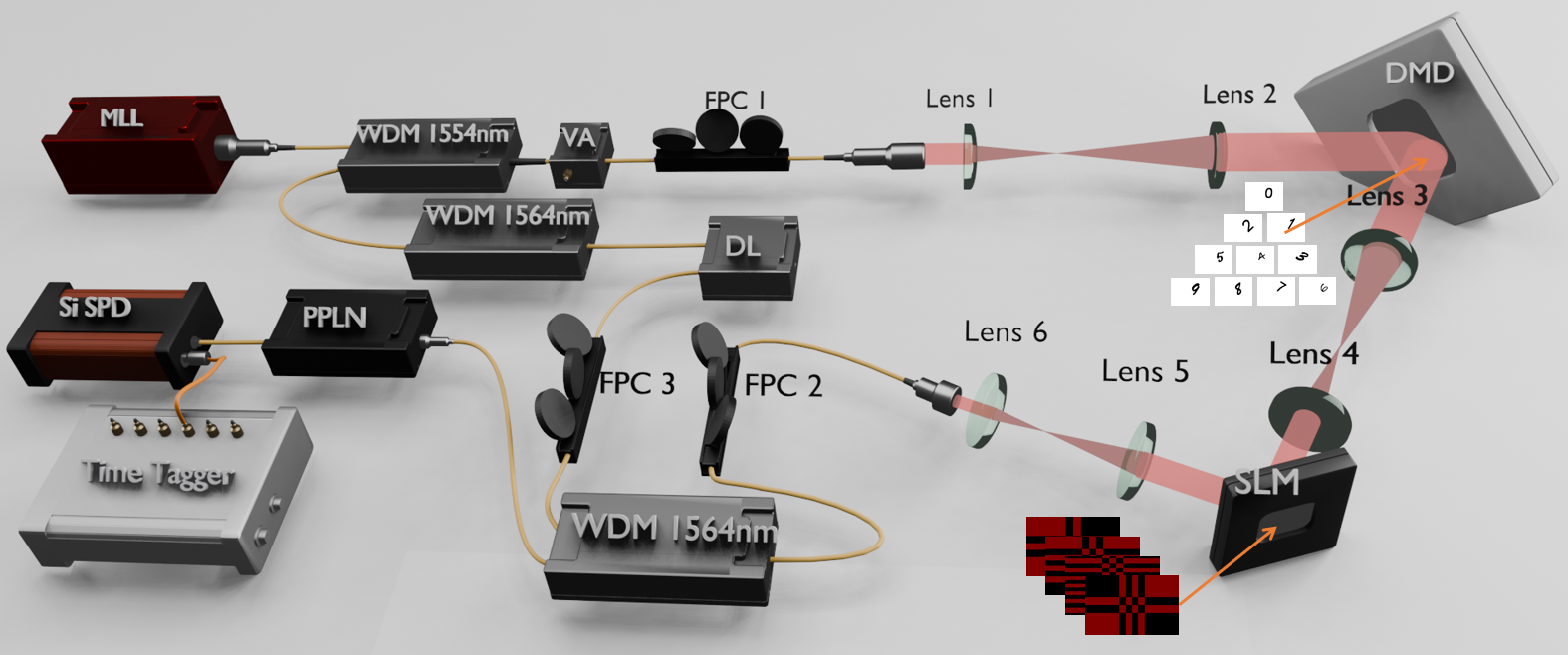} 

	\caption{\textbf{The experimental setup for first photon machine learning.}
		MLL, mode locked laser; SLM, spatial light modulator; PPLN, magnesium-doped periodic poled lithium niobate crystal; DMD, digital micromirror device; DL, delay line; VA, variable attenuator; FPC, fiber polarization controller; SPD, single photon detector.}
	\label{Fig.1} 
\end{figure}

Figure \ref{Fig.1} outlines the experimental setup, where the encoding is realized in reflective instead of transmissive optics. It begins with a femtosecond mode-locked laser (MLL, CALMAR LASER, FPL-03CFF). Wavelength division multiplexers (WDMs) with 200 GHz bandwidth are employed as spectral filters to generate the probe (1554 nm) and pump (1564 nm) pulse trains. A fiber polarization controller (FPC) is used to adjust the polarization of the probe light. An electrically controlled variable attenuator (VA, V1550A, Thorlabs) is applied next to precisely tune its probe power at a nanowatt level. Afterwards, it is collimated and launched into a free-space setup consisting of a digital micromirror device (DMD, Texas Instruments), three 4-F imaging lens systems, and a liquid crystal spatial light modulator (SLM, SLM210, Santec). The DMD is used to encode the MNIST patterns onto the photons, while the SLM is used to imprint onto the photons 10 phase masks, each for one of the 10 labels. After the DMD and SLM, the signal photons are coupled into a single-mode fiber (SMF), whose coupling efficiency is about 40\% before any modulation is applied either by the DMD or SLM. Then, the photons are combined with the pump via a WDM and sent into a magnesium-doped periodically poled lithium niobate (PPLN, HC Photonics) waveguide for frequency upconversion quantum parametric mode sorting \cite{Shahverdi2017,Rehain2020,PhysRev.125.475,PhysRevX.5.041017}. To achieve high efficiency, the polarization of the probe and pump is each controlled by a FPC. The upconverted signal is detected by a silicon single-photon detector (Excelitas, SPCM-AQRH-12-FC), and the time tags are recorded by a time tagger (Time Tagger Ultra, Swabian Instruments). Further information on those experimental components can be found in \cite{Li2024,Li:24}. This setup combines spatial light modulation for high dimensional information coding and quantum parametric mode sorting (QPMS) to ensure mode-projective measurement with high purity.  

The experimental procedure is as follows. First, a MNIST digit pattern is displayed on the DMD. Then, the 10 phase masks are sequentially uploaded on the SLM screen while the reflected photons are collected in the single mode fiber and detected with QPMS. The rotation stops when the first photon detection is registered, and the pattern is assigned to the corresponding label. Limited mainly by the low refreshing rate of the SLM (about 0.7 frames per second), the total time to complete the measurement for each SLM mask is about 1.4 seconds, so that it takes at least 14 seconds to finish each digit pattern (note actual time needs to be much longer; see explanations in the next paragraph). Then, the DMD mask is refreshed to display the next digit pattern, and the measurement process repeats. The procedure continues till all digits are measured. If the photon counts are statistically significant, another run will repeat for all digits.   

During the process, the average photon detection rate during the display of each SLM mask must be much less than 1, to minimized the bias arising from the sequence of mask displaying and detector dead time. In experiment, we first upload an all-black pattern on the DMD and a uniform phase mask onto the SLM with a $2\pi$ phase value for each pixel. Then, we use the VA to adjust the probe power level so that at most 0.1 photon is detected per pulse. For the MNIST digits, the detected photon number per pulse is between 0.001 and 0.06 on average.

\begin{figure}[htp]
    \centering
    {\includegraphics[width=0.49\linewidth]{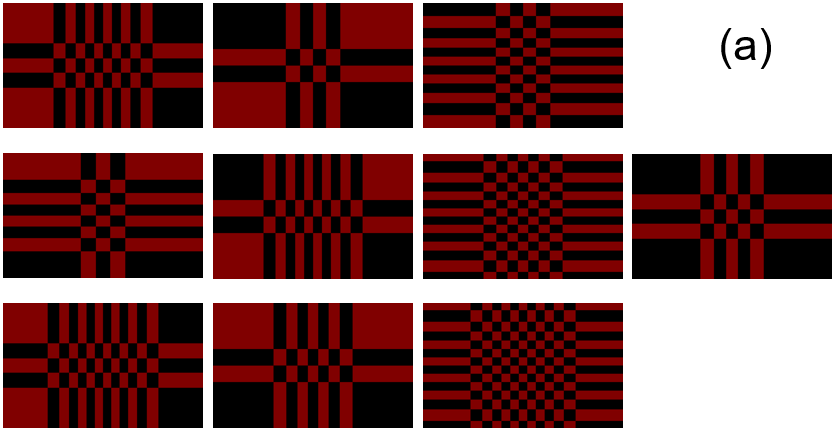}}
    {\includegraphics[width=
    0.49\linewidth]{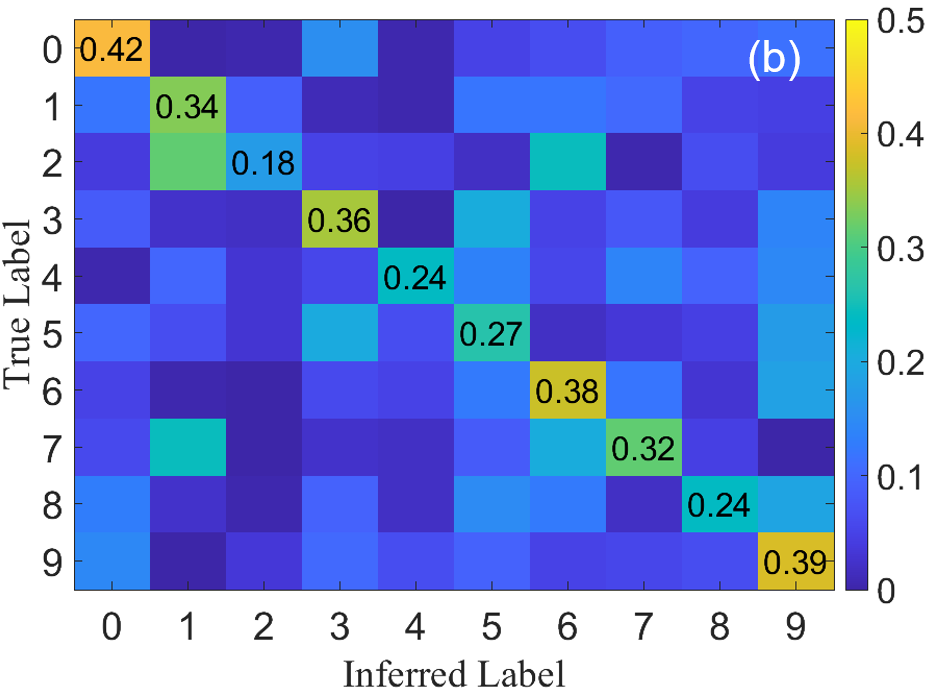}}
\caption{\textbf{Experimental QC}, where (\textbf{A}) are the phase masks chosen for Label 0 to 9 (ordering from left to right and top to down), and (\textbf{B}) is the confusion matrix, with an average fidelity of $31.00\%\pm0.77\%$.}
\label{Fig.Cp}
\end{figure}

In our experiment, we consider the first 100 hand-written patterns of each digit in the MNIST library, so that the total pattern number is 1000. For each label, we choose a Hermite-Gaussian (HG) mode such that when it is on display and a photon is detected, the Bayesian probability of the digit pattern under test in this label is the highest among all labels. There are, of course, other feasible mode selections, and one may use optimization methods to find them\cite{Xue2024}. In this work, however, we restrict our choices to a dozen of Hermite Gaussian modes, in hope to show that even without much optimization, the quantum effects can give superior performance beyond what an ideal classical system can theoretically achieve. For the 1000 patterns under test, we first select HG$_{11,4}$, HG$_{5,3}$, HG$_{5,11}$, HG$_{4,7}$, HG$_{11,3}$, HG$_{8,13}$, HG$_{6,4}$, H$_{13,4}$, HG$_{8,3}$, and HG$_{12,13}$ for Label 0 to 9, respectively, as shown in figure \ref{Fig.Cp}(a).

The experimental results are shown in Fig.~\ref{Fig.Cp}(b), where two rounds of measurements are carried out. For each pattern of Label $j$, the first-photon detection events are registered along with the corresponding phase mask $k$. 
Then, for each label, those detection events are aggregated, and the probability of inferring a digit of true Label $j$ as Label $k$, $P_{j\rightarrow k}$, is calculated by their normalization within the label class. Figure~\ref{Fig.Cp}(b) plots the confusion matrix of $\{P_{j\rightarrow k}\}$, where the diagonal numbers give the probability of correct classification, i.e., the probability for a digit pattern of true Label $j$ being inferred as Label $j$. For all labels, the success probability is well above 0.1 of random guess. The highest fidelity is 42\%, achieved for Label ``0'', and lowest is 18\% for Label 2. Except for Label 2, all labels have the highest probability of being corrected classified. The average fidelity for all labels in $31.00\% \pm 0.77\%$, where the uncertainty is calculated based on the shot-noise statistics of photon counting.  

\begin{figure}
    \centering
    {\includegraphics[width=0.49\linewidth]{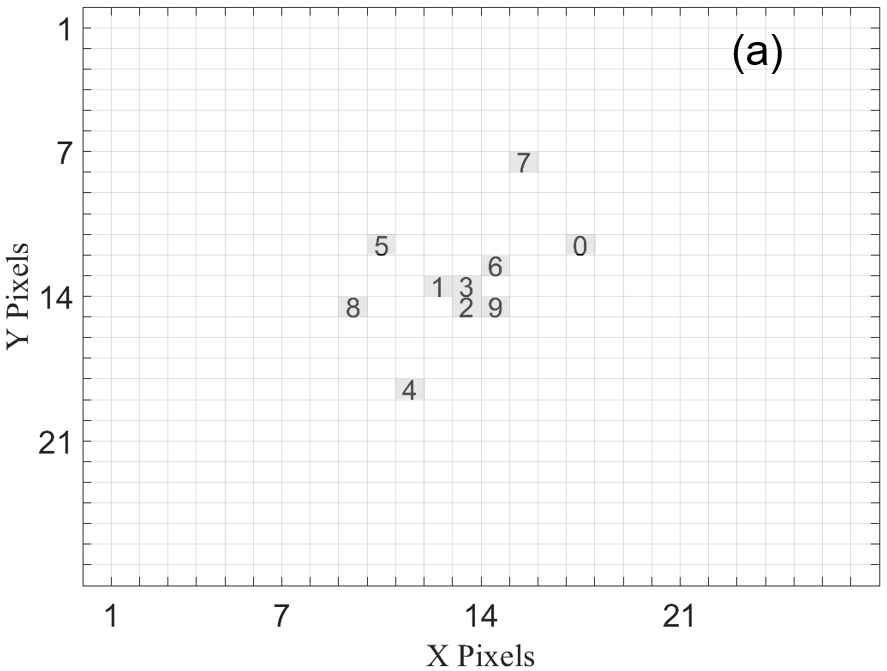}}
    {\includegraphics[width=
    0.49\linewidth]{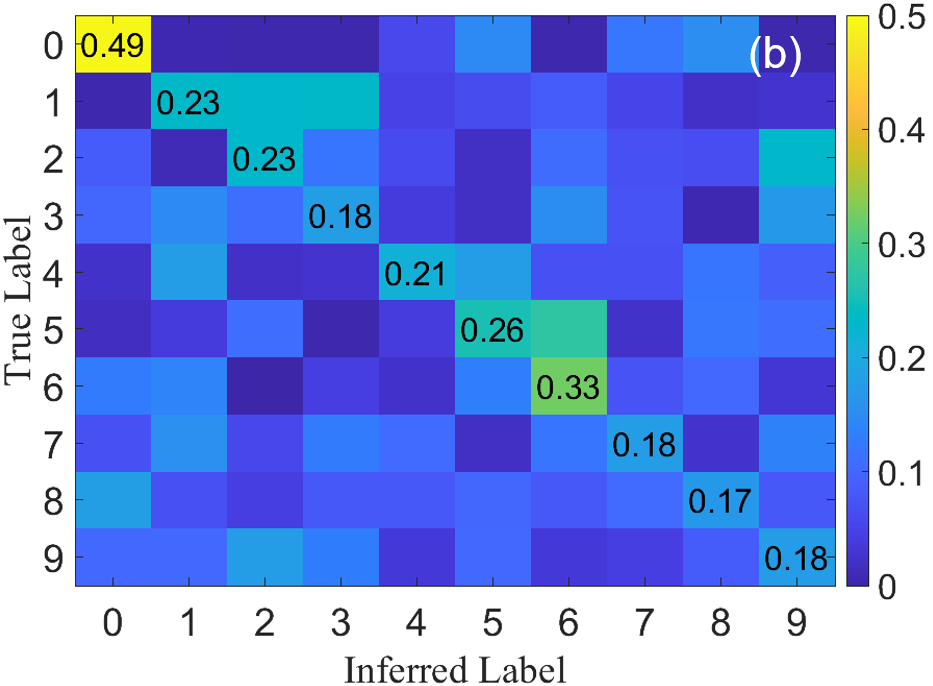}}
\caption{\textbf{Simulation results of an ideal CC}, (\textbf{A}) Pixel positions for each Label. (\textbf{B}) Confusion matrix, with a 24.57\% classification fidelity.}
\label{Fig.cls}
\end{figure}

As a comparison, Fig.~\ref{Fig.cls} shows the thought-experiment results for the ideal CC. Rather than projecting the encoded photons onto 10 HG modes, they are detected in 10 pixels that maximize the overall classification fidelity \cite{fischbacher2020single}. Specifically, for each label $j$, the index pixel $I_j$ is identified so that once a photon is detected in it, the Baysian probability for the photon to be encoded in Label $j$ is the highest among all 10 labels. If the same pixel is identified for multiple labels, then other pixels are chosen to give the next highest probabilities. Figure~\ref{Fig.cls}(a) shows the pixels chosen for the same MNIST patterns as in QC,  where for Label $0$ to $9$, their coordinates are (12,18), (14,13), (15,14), (14,14), (19,12), (12,11), (13,15), (8,16), (15,10), and (15,15), respectively. As seen, those pixels distribute around the center regime, where the information-carrying features are mostly located. 

To compute the fidelity, each digit pattern is tested by all index pixels ${I_1, I_2,...I_9}$, by checking which and how many of them are contained in the pattern. Then, we assign the pattern with equal probability to the labels those pixels indicate. For example, if a pattern contains $I_1,I_4,I_7$, then it has a $1/3$ probability to be inferred as of Label $1,4$ and $7$, respectively. In rare cases when a pattern contains no index pixel, then it will be assigned to any of the 10 labels, each with 1/10 probability. The classification accuracy for each label is then the statistical average for all patterns under it. Averaging the accuracies over all labels then gives the classification fidelity for CC. Figure~\ref{Fig.cls} lists the average accuracies for each label, where the highest accuracy of $49\%$ is achieved for Label ``0'', and the lowest of $17\%$ is achieved for Label ``8''. The average over all labels is $24.57\%$, close to the theoretical threshold for all pixels and all patterns \cite{fischbacher2020single,4426127}.

Comparing the results in Figs.~\ref{Fig.cls} and \ref{Fig.Cp}, QC clearly outperforms CC, whose classification fidelity of $31.00\%$ is significantly higher than $24.57\%$. This exhibits an unambiguous advantage originating from the quantum superposition principle. We note, however, that this advantage is only in the sense of statistical average. It is does not imply that the current QC is advantageous for every pattern or every class. In fact, in the present results, higher classification accuracies are obtained with the ideal CC for Label ``0''. This is because CC only samples a particular pixel of the patterns. For those with a highly distinguishing pixel, CC can tell them easily apart. In contrast, QC samples simultaneously all pixels; in fact, it probes the correlation among those pixels through interference measurement. As such, it is advantageous in capturing the wholistic features, similar to how visual information is processed in eyes.       

\begin{figure}
    \centering
    {\includegraphics[width=0.49\linewidth]{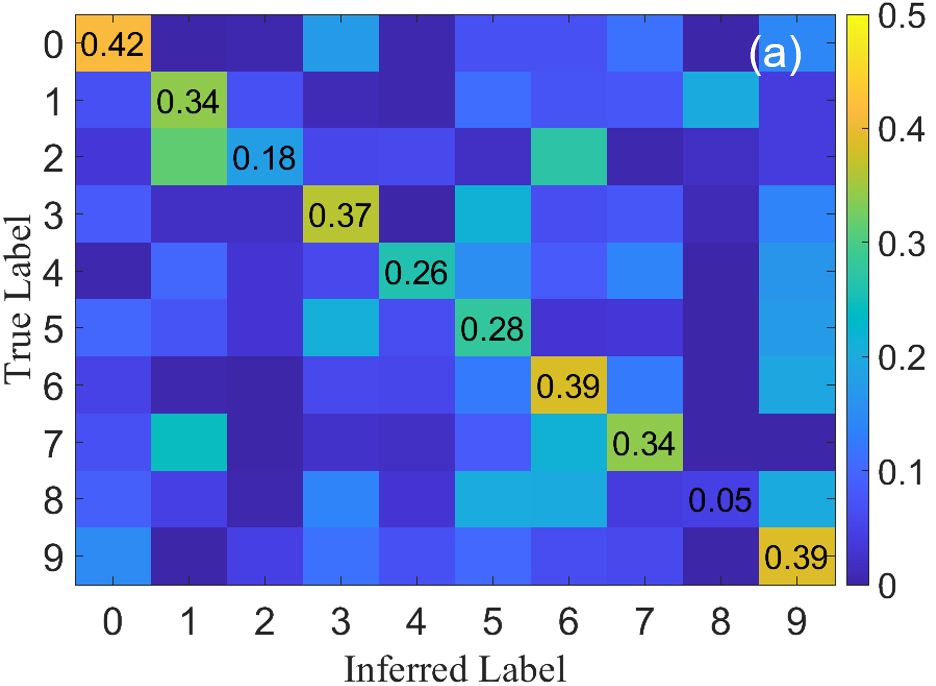}}
    {\includegraphics[width=
    0.49\linewidth]{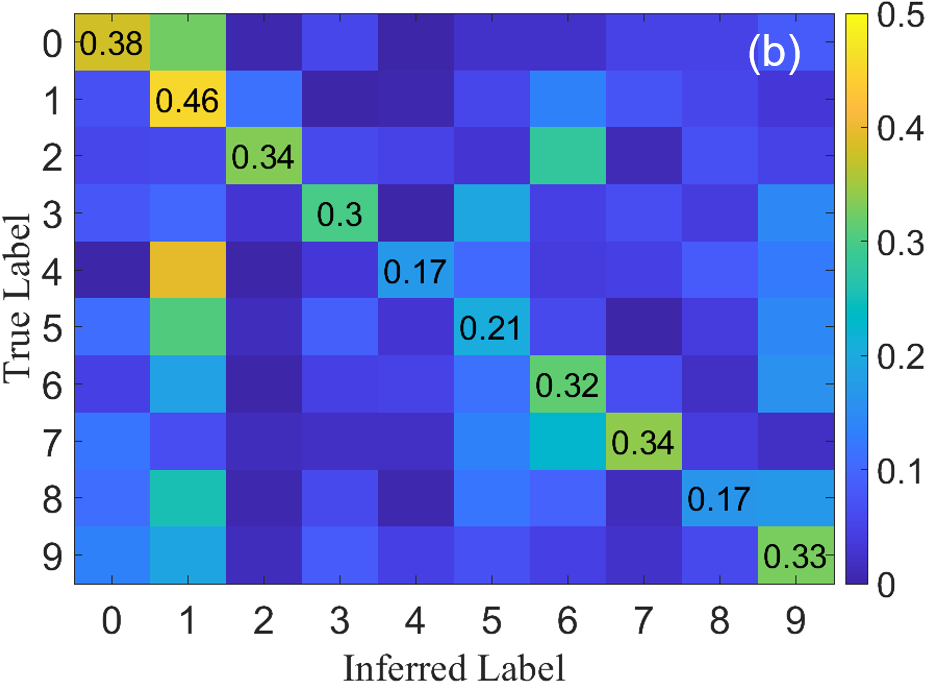}}
\caption{\textbf{Additional confusion matrices for QC with different phase mask sets}, whose classification fidelity is 30.00\%$\pm$1.14\% (\textbf{A})and 29.95\%$\pm$0.93\% (\textbf{B}).}
\label{Fig.HG}
\end{figure}

Finally, we would like to emphasize that the above comparison is made between the theoretical results of an ideal CC and experimental results of an under-optimized QC. The observed advantage is not specific to the particular patterns or phase masks we choose. Rather, it should be widely applicable and can provide significant benefits in many machine learning scenarios. As an example, in Fig.\ref{Fig.HG}, we show the fidelity matrice for two different sets of HG modes, where Fig.\ref{Fig.HG} (a) is for \{HG$_{11,4}$, HG$_{5,3}$, HG$_{5,11}$, HG$_{4,7}$, HG$_{11,3}$, HG$_{8,13}$, HG$_{6,4}$, H$_{13,4}$, HG$_{3,4}$, and HG$_{12,13}$\}, and Fig.\ref{Fig.HG} (b) is for \{HG$_{11,4}$, HG$_{5,4}$, HG$_{5,11}$, HG$_{4,7}$, HG$_{11,3}$, HG$_{8,13}$, HG$_{6,4}$, H$_{13,4}$, HG$_{8,3}$, and HG$_{12,13}$\}. The the classification fidelities are 30.00\%$\pm$1.14\% and 29.95\%$\pm$0.93\%, respectively, similar to Fig.\ref{Fig.Cp}.

\section{CONCLUSION}
In conclusion, we have demonstrated first photon machine learning, where the entire neural network for image classification is realized in optics and upon the detection of a single photon. Our experimental results find quantum effects to give a significant performance boost over what's possible with a purely classical neural network. This is the first unambiguous demonstration of quantum advantages---in this case offered by the quantum superposition principle---for AI. Also, the fact that all computations are realized on a single photon gives a new and perhaps the ultimate record of energy efficiency, which highlights a viable pathway to meet the fast growing power demands by the blooming AI industry. 

On the other hand, we would like to note that this advantage is demonstrated in a specifically-designed, narrowly-purposed setting. For other AI tasks, one would need to carefully design the architecture to harvest the power of quantum effects. We believe, however, that quantum mechanics can indeed help machine learning, and a hybrid quantum-classical system can establish practical advantages for realistic AI applications in near term. Finally, we would like to point out that although only a single photon is detected for decision making, the whole system energy consumption is much more than $10^{-19}$, when counting the power to run the optical and electronic devices and including that of emitted but undetected photons. To further reduce the total energy consumption will require additional replacement of electronics with optics. 

\clearpage 

%
\bibliography{science_template} 
\bibliographystyle{sciencemag}

%
%
%
%
%
%


\paragraph*{Funding:}
LL and SK were supported by the ACC-New Jersey under Contract No. W15QKN-18-D-0040.
\paragraph*{Author contributions:}
YH conceived the experiment with LL. LL designed the setup and conducted the experiment with help from others. LL, YH, and SK analyzed the results. All authors contributed to writing the manuscript.
\paragraph*{Competing interests:}
There are no competing interests to declare.
\paragraph*{Data and materials availability:}

The related original data (which is more than 100 GB) and codes will be uploaded on google drive, GitHub, or other website that allow such big file folder.

\end{document}